\documentclass[11pt,a4paper]{article}
\usepackage{graphicx,xcolor} 
\usepackage{jheppub}
\usepackage{amsmath}
\usepackage{amssymb}
\usepackage{mathtools}
\usepackage{comment}
\usepackage{tabularx}
\usepackage{booktabs}
\usepackage[utf8]{inputenc}
\usepackage{physics}
\usepackage{tikz}
\usetikzlibrary{calc,arrows.meta,decorations,decorations.pathmorphing,positioning,decorations.markings}
\allowdisplaybreaks

\makeatletter
\toccontinuoustrue
\makeatother

\title{All-multiplicity $\textrm{AdS}_4$ correlators in the self-dual sector}
\author[a,b]{Humberto Gomez,}
\author[b]{Renann Lipinski Jusinskas,}
\author[c]{Arthur Lipstein,}
\author[d]{Cristhiam Lopez-Arcos}

\affiliation[a]{Facultad de Ciencias Basicas,  Universidad Santiago de Cali,\\
Calle 5 $N^\circ$  62-00 Barrio Pampalinda, Cali, Valle, Colombia}
\affiliation[b]{Institute of Physics of the Czech Academy of Sciences \& CEICO \\ 
Na Slovance 2, 18221 Prague, Czech Republic}
\affiliation[c]{Department of Mathematical Sciences, Durham University, Durham, DH1 3LE, UK}
\affiliation[d]{Departamento de Matem\'{a}ticas, Universidad Nacional de Colombia Sede Medell\'{i}n,\\
Carrera 65 $\#$ 59A--110, Medell\'{i}n, Colombia}

\abstract{
It was recently shown that tree-level gluon correlators of any multiplicity in $\textrm{AdS}_4$ admit an expansion in terms of flat-space scattering amplitudes, where lower-point amplitudes in the expansion depend on merged external data. We derive here analogous formulae for correlators of scalar theories which encode the self-dual sectors of Yang-Mills theory and gravity in this background. In the latter case, the amplitudes are differential operators which can be obtained from the former using a simple double copy prescription in terms of kinematic numerators satisfying color-kinematics duality.
}

\begin{document}
\maketitle

\section{Introduction}

Self-dual Yang-Mills (SDYM) and self-dual gravity (SDG) are useful toy models for Yang-Mills theory and Einstein gravity in flat background since they can be described in terms of simple Lagrangians \cite{Plebanski:1975wn,Chalmers:1996rq} which make color-kinematics duality manifest \cite{Monteiro:2011pc,Boels:2013bi}, play a prominent role in twistor theory \cite{Penrose:1976js,Ward:1977ta}, and have very simple scattering amplitudes \cite{Bern:1993qk,Bern:1996ja,Bern:1998xc,Bern:1998sv} -- see \cite{Miller:2026fgp} for a pedagogical review. In recent years, there has been a great deal of interest in extending many of these results to non-trivial backgrounds \cite{Adamo:2017nia, Adamo:2020yzi,Adamo:2022mev,Neiman:2023bkq,Lipstein:2023pih,Chowdhury:2024dcy, Skvortsov:2026gtq,Skvortsov:2026ofl,Skvortsov:2026dru,Heuveline:2026knw,Bittleston:2023bzp,Bogna:2024gnt,Crawley:2023brz,Adamo:2023fbj,CarrilloGonzalez:2024sto}. In this paper, we will be primarily interested in the AdS$_4$ background since this is relevant for the study of AdS/CFT and cosmology (after Wick rotation to dS$_4$). In \cite{Lipstein:2023pih,Krasnov:2024qkh,Dunajski:2026dfe}, SDYM and SDG in AdS$_4$ were shown to admit a simple description in terms of scalar fields and boundary correlators of these scalar theories were subsequently studied in \cite{Chowdhury:2024dcy}. More recently, spinning SDYM correlators correlators were computed in \cite{Skvortsov:2026gtq,Skvortsov:2026ofl,Skvortsov:2026dru,Heuveline:2026knw}.

Our goal in this paper is to derive all-multiplicity formulae for boundary correlators of the scalar theories describing SDYM and SDG in AdS$_4$. Such a formula was recently obtained for full Yang-Mills theory in the form of an amplitude expansion \cite{Gomez:2026yno}: notably a sum over flat space scattering amplitudes  weighted by poles in the total energy. By energy, we refer to the sum of momenta for each leg of the correlator along the bulk direction of AdS$_4$, which is not conserved (the analogy with energy becomes more natural after Wick-rotating to dS$_4$ since the bulk coordinate becomes time-like). In the flat space limit of an $n$-point correlator, the total energy is taken to zero and the dominant term in the expansion is the $n$-point flat space amplitude \cite{Raju:2012zr}. The lower-point amplitudes appearing in the expansion are obtained by merging the external data of higher-point amplitudes. The primary tool used to derive this formula was Berends-Giele recursion \cite{Berends:1987me,Hasuwannakit:2025agr,Miller:2026fgp}, which was generalised to (A)dS background in \cite{Armstrong:2022mfr}. 

The advantage of this representation is that boundary correlators can be completely reconstructed to any multiplicity from flat space scattering amplitudes, which in principle allows for the direct extension of flat space techniques to curved background. On the other hand, the formula in \cite{Gomez:2026yno} was expressed in terms of general polarizations, which obscures properties such as the double copy to Einstein gravity. As we will see, SDYM and SDG provide a useful stepping stone in this regard since their correlators exhibit colour-kinematics duality and a double copy (see e.g. \cite{Correa:2024mub,Raeymaekers:2025akr}). For recent reviews of these concepts in flat and curved backgrounds, we refer the reader to \cite{Bern:2019prr} and \cite{CarrilloGonzalez:2026phk}, respectively. For curved backgrounds, we emphasize that these properties are made manifest by the amplitude expansion and were not previously understood in a systematic way.

This paper is organized as follows. In section \ref{review}, we review SDYM, SDG, and the amplitude expansion in AdS$_4$. In sections \ref{sec:SDYM-correlators} and \ref{SDGcorr} we then derive all-multiplicity formulae for scalar SDYM and SDG correlators, respectively. In section \ref{colorkin} we then explain how colour-kinematics duality is realised by the correlators. In section \ref{conclusion}, we end with some conclusions and future directions. 

\section{Review} \label{review}

In this section we will review the amplitude expansion of correlators in $\textrm{AdS}_4$, focusing on the $\textrm{Tr}(\phi^3)$ theory for simplicity, and then present a formulation of SDYM and SDG in this background.

\subsection{Amplitude expansion}

We will work in the Poincar\'e patch of Euclidean $\textrm{AdS}_4$ with metric 
\begin{equation}
ds^{2}=z^{-2} \eta_{mn}dx^{m}dx^{n}=z^{-2}( \eta_{\mu\nu}dx^{\mu}dx^{\nu}+dz^{2}),
\label{adsmetric}
\end{equation}
where $z \geq 0$ describes the radial direction, and $\eta_{\mu \nu}$ is the boundary metric. In practice we will perform a Weyl transformation to half of flat space with boundary at $z=0$. Due to the presence of the boundary, momentum is not conserved along the $z$ direction. However, we may still Fourier transform correlators along the boundary directions and assign a null four-momentum to each external leg of a correlator, i.e. $k_m = \{k_\mu, \mathrm{i} k\}$, with $k = \sqrt{k \cdot k}$, where $a\cdot b \equiv \eta^{\mu \nu} a_\mu b_\nu$. Each external leg is understood to be a single-particle state, labeled by a letter $p$. More generally, we will denote multi-particle labels (words) by capital letters, e.g. $P=P_1\ldots P_i$, where the labels $P_j$ denote sub-words. The boundary momentum for such a state is then given by $k^{\mu}_{P}= k^{\mu}_{P_1}+\ldots +k_{P_i}^{\mu}$ and its energy is
\begin{equation}
    E_{P_1,\ldots,P_i}=k_{P_1}+\ldots + k_{P_i}.
\end{equation}
with $k_P=\sqrt{k_P \cdot k_P}$. We refer to these as merged momenta, since the boundary components are sums of single-particle momenta and bulk components are defined such that the four-momentum is null. Moreover, it is convenient to introduce Lorentz-invariant Mandelstam variables in terms of (merged) momenta as follows:
\begin{equation}\label{eq:GeneralSij}
    S_{P_{1},\ldots,P_{i}} \equiv 
  k^2_{P_1\ldots P_i} - E_{P_1,\ldots,P_i}^2.
\end{equation}

In order to illustrate the amplitude expansion of boundary correlators in $\textrm{AdS}_4$, a useful toy model to consider is the $\textrm{Tr}(\phi^3)$ theory in half of flat space. Using conventional Feynman diagrams with Dirichlet boundary conditions on the bulk-to-bulk propagators, the color-ordered tree-level four-point correlator is given by \cite{Chowdhury:2024dcy,Gomez:2026yno} 
\begin{equation}
\mathcal{C}(1,2,3,4) = \frac{\phi_{1}\phi_{2}\phi_{3}\phi_{4}}{E_{1,2,3,4} E_\textrm{L} E_\textrm{R}}+(\{1,2,3\}\to \{2,3,1\}),\label{eq:4ptphi3-usual}
\end{equation}
where $E_\textrm{L}=E_{1,2,34}$ and $E_\textrm{R}=E_{12,3,4}$ and $\phi_i$ are the single-particle (scalar) polarisations. A key observation is that the denominator can be written as 
\begin{equation}\label{eq:identityE}
\frac{1}{E_{1,2,3,4} \, E_\textrm{L}\,  E_\textrm{R}} = \frac{1}{S_{1,2}}\left(\frac{1}{E_{1,2,3,4}} -\frac{1}{E_{12,3,4}} \right).
\end{equation}
where we see the emergence of flat space Mandelstam variables despite the lack of energy conservation. As a result, the four-point correlator can be recast as 
\begin{equation}
\mathcal{C}(1,2,3,4)=\frac{A (1,2,3|4)}{E_{1,2,3,4}} + \frac{A (12,3|4)}{E_{12,3,4}} +\frac{A (1,23|4)}{E_{1,23,4}}, \label{eq:4ptphi3-flat}
\end{equation}
where we recognize the first term as a four-point flat space amplitude times a four-point energy pole. The other two terms have three-point energy poles, and the respective three-point amplitudes are expressed in terms of merged momenta and merged polarisations defined by 
\begin{equation}
    \Phi^{(1)}(ij) = - \frac{1}{S_{i,j}}  \phi_i \phi_j. \label{eq:phi3-merdeg2pt}
\end{equation}

Remarkably, the decomposition in \eqref{eq:4ptphi3-flat} can be extended to any multiplicity. It is given by
\begin{equation}
  \mathcal{C}(1,\ldots,N,N+1)=\sum_{i=2}^{N} \, \sum_{1\cdots N=P_{1}\cdots P_{i}}\frac{A (P_1,\ldots,P_i|N+1)}{E_{P_{1},\ldots,P_{i},n}}.
\label{eq:phi3npt}
\end{equation}
where $A (P_1,\ldots,P_i|N+1)$ denote flat space (partial) amplitudes at $(i+1)$-points computed via the Berends--Giele prescription, and the merged polarisations entering the lower-point amplitudes can be obtained recursively. A similar representation exists for tree-level Yang-Mills correlators in axial gauge. For more details, see \cite{Gomez:2026yno,LipinskiJusinskas:2026ctz}.

\subsection{Self-dual theories}

We will now review the construction of scalar theories describing the self-dual sectors of Yang-Mills and Einstein gravity in $\textrm{AdS}_4$, which are the main focus of this paper. For this purpose, it is convenient to use light-cone coordinates
\begin{equation}
\begin{array}{ccccccc}
u=\mathrm{i}t+z, && v=\mathrm{i}t-z, && w=x+\mathrm{i}y, && \bar{w}=x-\mathrm{i}y.\end{array}
\end{equation}

The momenta in light-cone variables are then given by
\begin{equation}
\begin{array}{ccccccc}
k_u = \frac{\mathrm{i}}{2}(E-k_{t}) , && k_{v}=-\frac{\mathrm{i}}{2}(E+k_{t}), && k_{w}=\frac{1}{2}(k_{x}-\mathrm{i}k_{y}), && k_{\bar{w}}=\frac{1}{2}(k_{x}+\mathrm{i}k_{y}).\end{array}
\end{equation}
such that
\begin{equation}
 \mathrm{i}k_{\mu}x^{\mu}-Ez=\mathrm{i}(uk_{u}+vk_{v}+wk_{w}+\bar{w}k_{\bar{w}}).
\end{equation}

As shown in \cite{Lipstein:2023pih}, SDYM in $\textrm{AdS}_4$ can be encoded in a theory with two scalars described by the following Lagrangian: 
\begin{equation}\label{eq:SDYM-lagrangian}
\mathcal{L}_{\textrm{SDYM}} = \textrm{Tr} \{ \bar\Phi ( \Box_{\mathbb{R}^{4}} \Phi + \mathrm{i} [\partial_u \Phi, \partial_w \Phi] )\},
\end{equation}
in which we neglect boundary terms at $z=0$. This is same Lagrangian found in flat space \cite{Chalmers:1996rq} due to the fact that (self-dual) Yang-Mills is classically scale-invariant. The Lagrangian can alternatively be written as
\begin{equation} 
\mathcal{L}_{\textrm{SDYM}}=\textrm{Tr}\left(\bar{\Phi}\Box\Phi+\mathrm{i}\bar{\Phi}\left[\left\{ \Phi,\Phi\right\} \right]\right), \label{LSDYM}
\end{equation}
where the Poisson bracket is given by
\begin{equation}
 \left\{ f,g\right\} =\partial_{u}f\partial_{w}g-\partial_{w}f\partial_{u}g,
\end{equation}
and satisfies a Jacobi identity analogous to that of the square brackets for the color commutator. This can be interpreted as a kinematic analogue of the Jacobi identity and reflects an underlying property of the SDYM action known as color/kinematics duality \cite{Monteiro:2011pc}.

The correlators of this theory were first analyzed in \cite{Chowdhury:2024dcy} using Dirichlet boundary conditions on the bulk-to-bulk propagators. The three-point and four-point correlators are given by 
\begin{align}
\mathcal{C}\left(1^{+},2^{+},3^{-}\right) &= \frac{X_{1|2} }{E_{1,2,3}},\label{eq:SDYM-3pt}\\
\mathcal{C}(1^{+},2^{+},3^{+},4^{-})&=\frac{X_{1|2}X_{3|4}}{E_{1,2,3,4}E_{12,3,4}E_{1,2,34}}+\frac{X_{2|3}X_{4|1}}{E_{1,2,3,4}E_{23,4,1}E_{2,3,41}}. \label{eq:SDYM-4pt}
\end{align}We denote by $+$ ($-$) the external legs associated to $\Phi$ ($\bar{\Phi}$), and $X_{i|j}$ can be seen as kinematic structure constants that also play a role in scattering amplitudes of SDYM in flat space \cite{Monteiro:2011pc}:
\begin{equation}
   X_{i|j}\equiv\mathrm{i}(k_{iw}k_{ju}-k_{iu}k_{jw}).
\end{equation}
Comparing \eqref{eq:SDYM-4pt} to the $\textrm{Tr}(\phi^3)$ correlator in \eqref{eq:4ptphi3-flat}, we see that similar manipulations can be applied to the four-point SDYM correlator in terms of four-point amplitude and a three-point amplitude in flat space (without energy conservation). We will describe the generalisation to arbitrary multiplicity in section \ref{sec:SDYM-correlators}.

Next let us turn to SDG in $\textrm{AdS}_4$. In this case, we have the following scalar Lagrangian \cite{Lipstein:2023pih}:
\begin{equation}\label{LSDG}
\mathcal{L}_{SDG} = \frac{\bar{\varphi}}{u-v}\Box_{\mathbb{R}^{4}}(\frac{\varphi}{u-v})-\bar{\varphi}\{\{\frac{\varphi}{u-v},\frac{\varphi}{u-v}\}\}_{*}\,,
\end{equation}
where the deformed nested Poisson bracket is given by
\begin{multline}
\left\{ \left\{ f,g\right\} \right\} _{*} = \frac{1}{2} \Big[\partial_u^2 f\ \partial_w^2 g + \partial_u^2 g \ \partial_w^2 f\Big] - \partial_u \partial_w f \ \partial_u \partial_w g\\
+ \frac{1}{u - v} \Big[ \partial_u\partial_w f\ \partial_w g + \partial_w f\ \partial_u \partial_w g  - ( \partial_w^2 f \ \partial_u g + \partial_u f \ \partial_w^2 g )   \Big].
\end{multline}
In the flat flat space limit $z \rightarrow \infty$, the deformation vanishes and the equations of motion reduce to those found in flat space \cite{Plebanski:1975wn,Monteiro:2011pc}. Note that we can obtain \eqref{LSDG} from \eqref{LSDYM} by the following replacements:
\begin{equation}
\Phi \rightarrow \frac{\varphi}{u-v},\,\,\,\ \frac{\mathrm{i}}{2}\left[\left\{ ,\right\} \right]\rightarrow\left\{ \left\{ ,\right\} \right\} _{*}.
\label{doublecopy}
\end{equation}
In other words, we can obtain the SDG action from the SDYM one by replacing the color bracket with a deformed Poisson bracket defined as
\begin{equation}
\left\{ f,g\right\} _{*}=\left\{ f,g\right\} +\frac{2}{u-v}\left(f\partial_{w}g-g\partial_{w}f\right).
\end{equation}
The deformed Poisson bracket also satisfies a kinematic Jacobi identity and the replacements in \eqref{doublecopy} can be interpreted as a double copy (for more details, see \cite{Lipstein:2023pih}).

As shown in \cite{Chowdhury:2024dcy}, the correlators in this theory are naturally expressed in terms of differential operators. At three points, for example, we have 
\begin{equation}
\mathcal{M}\left(1^{+},2^{+},3^{-}\right)=X_{1|2}\mathcal{D}_{1|2}\left(\frac{1}{E_{1,2,3}}\right)
\end{equation}
where  $\mathcal{D}_{1|2}= k_{2w}-k_{1w}-X_{1|2}\partial_{E_{1,2}}$. It is also possible to express higher-point correlators in terms of differential operators acting on $\phi^3$ correlators in half of flat space, although a systematic procedure for doing so was not presented in \cite{Chowdhury:2024dcy}. In section \ref{SDGcorr}, we will present an all-multiplicity formula in the form of a differential amplitude expansion, where the amplitudes are obtained from those of SDYM by inserting differential operators in the interaction vertices. This has a natural interpretation in terms of the double copy, which is made manifest by the Lagrangians above. 

\section{Correlators in self-dual Yang--Mills}\label{sec:SDYM-correlators}

We compute the color ordered correlators of the self-dual Yang--Mills theory via classical multi-particle solutions of
\begin{equation}
    \Box_{\mathbb{R}^{4}}\Phi+\mathrm{i}[\partial_{u}\Phi,\partial_{w}\Phi]=0.\label{eq:SDYM-eom}
\end{equation}
This is the equation of motion derived from the Lagrangian \eqref{eq:SDYM-lagrangian}, with $\Box_{\mathbb{R}^{4}}$ denoting the flat-space d'Alembertian. The single-particle solutions of \eqref{eq:SDYM-eom} are given by $\Phi(z,k)=\exp (\mathrm{i}k_{p\mu}x^{\mu}-E_{p}z)$, with $\mu=t,x,y$. The respective multi-particle solutions can be written in a color-stripped form as
\begin{align}
\Phi(x,z) & = \sum_{P}T^{a_{P}}e^{\mathrm{i}k_{P\mu}x^{\mu}} \Phi_{P}(z),\\
\Phi_{P}(z) &=\sum_{i=1}^{|P|} \sum_{P=P_{1}\ldots P_{i}} \Phi^{(i)}(P_{1},\ldots,P_{i})e^{-E_{P_{1},\ldots,P_{i}}z},
\end{align}
where the word $P=p_1 \ldots p_n$ has length $|P|=n$, and  $T^{a_{P}} = T^{a_{p_1}} \ldots T^{a_{p_n}}$ is a product of gauge group generators. The multi-particle coefficients $\Phi^{(i)}(P_{1},\ldots,P_{i})$ are recursively determined via the equation of motion, with $P=P_1 \ldots P_i$, and expressed as
\begin{equation}
    S_{P_{1},\ldots,P_{i}}\Phi^{(i)}(P_{1},\ldots,P_{i}) = \sum_{j=1}^{i-1}X_{P_{1},\ldots,P_{j}|P_{j+1},\ldots,P_{i}}\Phi^{(j)}(P_{1},\ldots,P_{j})\Phi^{(i-j)}(P_{j+1},\ldots,P_{i}),
\end{equation}
with
\begin{equation}
\label{Xs}
X_{P_{1},\ldots,P_{j}|P_{j+1},\ldots,P_{i}}\equiv\mathrm{i}[(k_{P_{1},\ldots,P_{j}})_{w}(k_{P_{j+1},\ldots,P_{i}})_{u}-(k_{P_{1},\ldots,P_{j}})_{u}(k_{P_{j+1},\ldots,P_{i}})_{w}].
\end{equation}
Note that $\Phi^{(1)}(P)$ parametrizes a free solution of equation \eqref{eq:SDYM-eom}, therefore it cannot be determined via the recursion above ($S_P=0$). It is fixed instead by demanding the homogeneous Dirichlet boundary condition (i.e. $\Phi_P(0)=0$ for $|P|>1$),
\begin{equation}
    \Phi^{(1)}(P)=-\sum_{i=2}^{|P|}\sum_{P=P_{1}\cup\ldots\cup P_{i}}\Phi^{(i)}(P_{1},\ldots,P_{i}).
\end{equation} For one-letter words, $\Phi^{(1)}(p)$ is set to $1$, corresponding to the external legs of the correlator.

Let us look at some examples to showcase the recursion. The current $\Phi^{(2)}(P_1,P_2)$ is simply given by
\begin{equation}
    \Phi^{(2)}(P_1,P_2)=\Phi^{(1)}(P_1)\Phi^{(1)}(P_2)\frac{X_{P_1|P_2}}{S_{P_1,P_2}}.\label{eq:SDYM-2current}
\end{equation} In this case, the merged (two-particle) polarization $\Phi^{(1)}(ij)$ is expressed as
\begin{equation}
    \Phi^{(1)}(ij) = -\frac{X_{i|j}}{S_{i,j}},
\end{equation}
which is the SDYM analogue of equation \eqref{eq:phi3-merdeg2pt}. For the current $\Phi^{(3)}(P_1,P_2,P_3)$, the recursion leads to
\begin{multline}
    \Phi^{(3)}(P_{1},P_{2},P_{3})=\frac{X_{P_{1},P_{2}|P_{3}}}{S_{P_{1},P_{2},P_{3}}}\Phi^{(2)}(P_{1},P_{2})\Phi^{(1)}(P_{3})+\frac{X_{P_{1}|P_{2},P_{3}}}{S_{P_{1},P_{2},P_{3}}}\Phi^{(1)}(P_{1})\Phi^{(2)}(P_{2},P_{3}), \\
    =\left(\frac{X_{P_{1},P_{2}|P_{3}}X_{P_{1}|P_{2}}}{S_{P_{1},P_{2},P_{3}}S_{P_{1},P_{2}}}+\frac{X_{P_{1}|P_{2},P_{3}}X_{P_{2}|P_{3}}}{S_{P_{1},P_{2},P_{3}}S_{P_{2},P_{3}}}\right)\Phi^{(1)}(P_{1})\Phi^{(1)}(P_{2})\Phi^{(1)}(P_{3}).\label{eq:SDYM-3current}
\end{multline}

Finally, color ordered correlators are defined as
\begin{equation}
    \mathcal{C}(1^{+},\ldots,N^{+},(N+1)^{-})=\sum_{i=2}^{N}\sum_{1\ldots N=P_{1}\ldots P_{i}}\frac{A(P_1,\ldots,P_i|N+1)}{E_{P_{1},\ldots,P_{i},N+1}},\label{eq:SDYM-Nptcorrelator}
\end{equation}
with
\begin{equation}
    A(P_1,\ldots,P_i|N+1) = S_{P_{1},\ldots,P_{i}}\Phi^{(i)}(P_{1},\ldots,P_{i}),\label{eq:SDYM-flat-amplitude}
\end{equation}
where particles $1$ to $N$ are associated to $\Phi$, and particle $(N+1)$ to $\bar{\Phi}$. The object $A(P_1,\ldots,P_i|N+1)$ can be interpreted as flat space (partial) amplitudes with generalized external states $\Phi^{(1)}(P_j)$ with positive helicity, keeping in mind that momentum is not conserved along the $z$ direction. Since four-point or higher flat space amplitudes vanish in the SDYM theory, higher point energy residues ($E_{P_{1},\ldots,P_{i},N+1} \to 0$, for $i>2$) are all zero, consistent with the flat space limit of the correlators.

Reproducing the three-point correlator \eqref{eq:SDYM-3pt} is a trivial task. The four-point correlator is initially expressed as
\begin{equation}
\mathcal{C}(1^{+},2^{+},3^{+},4^{-})=\frac{1}{E_{1,2,3,4}}A(1,2,3|4)
+\frac{1}{E_{12,3,4}}A(12,3|4)+\frac{1}{E_{1,23,4}}A(1,23|4),
\end{equation}
where we have
\begin{align}
    A(1,2,3|4)&=\frac{X_{1,2|3}X_{1|2}}{S_{1,2}}+\frac{X_{1|2,3}X_{2|3}}{S_{2,3}},\\A(12,3|4)&=-\frac{X_{12|3}X_{1|2}}{S_{1,2}},\\A(1,23|4)&=-\frac{X_{1|23}X_{2|3}}{S_{2,3}}.
\end{align}
These expressions are directly obtained from equations \eqref{eq:SDYM-2current}, \eqref{eq:SDYM-3current}, and \eqref{eq:SDYM-flat-amplitude}.
In order to get to the simple form presented in equation \eqref{eq:SDYM-4pt}, we simply use
\begin{align}
    X_{1,2|3}&=X_{3|4}+\frac{1}{2}E_{1,2,3,4}(k_{3})_{w},\\X_{12|3}&=X_{3|4}+\frac{1}{2}E_{12,3,4}(k_{3})_{w},\\X_{1|2,3}&=X_{4|1}-\frac{1}{2}E_{1,2,3,4}(k_{1})_{w},\\X_{1|23}&=X_{4|1}-\frac{1}{2}E_{1,23,4}(k_{1})_{w},
\end{align}
which follow from momentum conservation along the boundary directions, and the identity displayed in equation \eqref{eq:identityE}.

The five-point correlator is given by
\begin{multline}
    \mathcal{C}(1^{+},2^{+},3^{+},4^{+},5^{-})=\frac{1}{E_{1,2,3,4,5}}A(1,2,3,4|5)\\+\frac{1}{E_{12,3,4,5}}A(12,3,4|5)+\frac{1}{E_{1,23,4,5}}A(1,23,4|5)+\frac{1}{E_{1,2,34,5}}A(1,2,34|5)\\+\frac{1}{E_{123,4,5}}A(123,4|5)+\frac{1}{E_{12,34,5}}A(12,34|5)+\frac{1}{E_{1,234,5}}A(1,234|5).
\end{multline}
The coefficients of the energy poles can be written as
\begin{align}
    A(1,2,3,4|5)&=\frac{X_{1,2,3|4}X_{1,2|3}X_{1|2}}{S_{1,2,3}S_{1,2}}+\frac{X_{1,2,3|4}X_{1|2,3}X_{2|3}}{S_{1,2,3}S_{2,3}}\nonumber\\&+\frac{X_{1|2,3,4}X_{2,3|4}X_{2|3}}{S_{2,3,4}S_{2,3}}+\frac{X_{1|2,3,4}X_{2|3,4}X_{3|4}}{S_{2,3,4}S_{3,4}}+\frac{X_{1,2|3,4}X_{1|2}X_{3|4}}{S_{1,2}S_{3,4}},\\A(12,3,4|5)&=-\frac{X_{12,3|4}X_{12|3}X_{1|2}}{S_{12,3}S_{1,2}}-\frac{X_{12|3,4}X_{1|2}X_{3|4}}{S_{1,2}S_{3,4}},\\A(123,4|5)&=\frac{X_{123|4}X_{12|3}X_{1|2}}{S_{12,3}S_{1,2}}-\frac{X_{123|4}X_{1,2|3}X_{1|2}}{S_{1,2,3}S_{1,2}}\nonumber\\&+\frac{X_{123|4}X_{1|23}X_{2|3}}{S_{1,23}S_{2,3}}-\frac{X_{123|4}X_{1|2,3}X_{2|3}}{S_{1,2,3}S_{2,3}},\\
    A(12,34|5)&=\frac{X_{12|34}X_{1|2}X_{3|4}}{S_{1,2}S_{3,4}},
\end{align}
with similar expressions for the other orderings.

Our construction makes the different diagram contributions straightforward to identify. For example,
\begin{eqnarray}
\begin{tikzpicture}[x=0.55cm,y=0.55cm, line cap=round, line join=round]
  \fill[gray!13] (0,0) circle (2.15);
    \draw[line width=0.2pt] (0,0) circle (2.15);

  \coordinate (p1) at (-1.95,0.98);
  \coordinate (p2) at (-1.95,-0.98);
  \coordinate (p5) at (0,2.15);
  \coordinate (p3) at (1.95,0.98);
  \coordinate (p4) at (1.95,-0.98);

  \coordinate (x1) at (-0.92,0);
  \coordinate (x2) at (0,0);
  \coordinate (x3) at (0.92,0);

  \draw[line width=0.6pt] (p1) -- (x1);
  \draw[line width=0.6pt] (p2) -- (x1);
  \draw[line width=0.6pt] (x1) -- (x2) -- (x3);
  \draw[line width=0.6pt] (p5) -- (x2);
  \draw[line width=0.6pt] (p3) -- (x3);
  \draw[line width=0.6pt] (p4) -- (x3);

  \fill (p1) circle (1.25pt);
  \fill (p2) circle (1.25pt);
  \fill (p3) circle (1.25pt);
  \fill (p4) circle (1.25pt);
  \fill (p5) circle (1.25pt);
  \fill (x1) circle (1.05pt);
  \fill (x2) circle (1.05pt);
  \fill (x3) circle (1.05pt);

  \node[font=\scriptsize, above left] at (-2.12,1.05) {$2^+$};
  \node[font=\scriptsize, below left] at (-2.12,-1.05) {$1^+$};
  \node[font=\scriptsize, above] at (0,2.34) {$3^+$};
  \node[font=\scriptsize, above right] at (2.12,1.05) {$4^+$};
  \node[font=\scriptsize, below right] at (2.12,-1.05) {$5^-$};

  \node[font=\scriptsize, below] at (-0.92,-0.18) {$\,$};
  \node[font=\scriptsize, below] at (0,-0.18) {$\,$};
  \node[font=\scriptsize, below] at (0.92,-0.18) {$\,$};

  \node[font=\scriptsize, above] at (-0.46,0.12) {$\,$};
  \node[font=\scriptsize, above] at (0.46,0.12) {$\,$};
\end{tikzpicture}
\qquad
\makebox[275pt][r]{\raisebox{6.6ex}{$\,\, =\, \frac{X_{1|2}}{S_{1,2}} \left(\frac{ X_{1,2|3}X_{1,2,3|4}}{S_{1,2,3}E_{1,2,3,4,5}} - \frac{X_{12|3}X_{12,3|4}}{S_{12,3}E_{12,3,4,5}} - \frac{X_{1,2|3}X_{123|4}}{S_{1,2,3}E_{123,4,5}}+ \frac{X_{12|3}X_{123|4}}{S_{12,3}E_{123,4,5}}\right), $}} \label{eq:SDYM-5pt-12345}
\end{eqnarray}
and
\begin{eqnarray}
\begin{tikzpicture}[x=0.55cm,y=0.55cm, line cap=round, line join=round]
  \fill[gray!13] (0,0) circle (2.15);
    \draw[line width=0.2pt] (0,0) circle (2.15);

  \coordinate (p1) at (-1.95,0.98);
  \coordinate (p2) at (-1.95,-0.98);
  \coordinate (p5) at (0,2.15);
  \coordinate (p3) at (1.95,0.98);
  \coordinate (p4) at (1.95,-0.98);

  \coordinate (x1) at (-0.92,0);
  \coordinate (x2) at (0,0);
  \coordinate (x3) at (0.92,0);

  \draw[line width=0.6pt] (p1) -- (x1);
  \draw[line width=0.6pt] (p2) -- (x1);
  \draw[line width=0.6pt] (x1) -- (x2) -- (x3);
  \draw[line width=0.6pt] (p5) -- (x2);
  \draw[line width=0.6pt] (p3) -- (x3);
  \draw[line width=0.6pt] (p4) -- (x3);

  \fill (p1) circle (1.25pt);
  \fill (p2) circle (1.25pt);
  \fill (p3) circle (1.25pt);
  \fill (p4) circle (1.25pt);
  \fill (p5) circle (1.25pt);
  \fill (x1) circle (1.05pt);
  \fill (x2) circle (1.05pt);
  \fill (x3) circle (1.05pt);

  \node[font=\scriptsize, above left] at (-2.12,1.05) {$4^+$};
  \node[font=\scriptsize, below left] at (-2.12,-1.05) {$3^+$};
  \node[font=\scriptsize, above] at (0,2.34) {$5^-$};
  \node[font=\scriptsize, above right] at (2.12,1.05) {$1^+$};
  \node[font=\scriptsize, below right] at (2.12,-1.05) {$2^+$};

  \node[font=\scriptsize, below] at (-0.92,-0.18) {$\,$};
  \node[font=\scriptsize, below] at (0,-0.18) {$\,$};
  \node[font=\scriptsize, below] at (0.92,-0.18) {$\,$};

  \node[font=\scriptsize, above] at (-0.46,0.12) {$\,$};
  \node[font=\scriptsize, above] at (0.46,0.12) {$\,$};
\end{tikzpicture}
\qquad
\makebox[215pt][r]{\raisebox{6.6ex}{$\,\, =\, \frac{X_{1|2}X_{3|4}}{S_{1,2}S_{3,4}} \left(\frac{ X_{1,2|3,4}}{E_{1,2,3,4,5}} - \frac{X_{12|3,4}}{E_{12,3,4,5}} - \frac{X_{1,2|34}}{E_{1,2,34,5}}+ \frac{X_{12|34}}{E_{12,34,5}}\right). $}}\label{eq:SDYM-5pt-34512}
\end{eqnarray}
We have also performed extensive numerical checks against the literature. In particular, we find agreement with the four-point and five-point correlators presented in \cite{Chowdhury:2024dcy}.

\section{Correlators in self-dual gravity} \label{SDGcorr}

After defining $\phi=\frac{\varphi}{2(u-v)}$, the equation of motion following from \eqref{LSDG} is given by
\begin{equation}
\Box_{\mathbb{R}^{4}}\phi = \frac{(u-v)}{2}(\partial_{w}^{2}\phi\partial_{u}^{2}\phi-(\partial_{u}\partial_{w}\phi)^{2})+(\partial_{w}\phi\partial_{u}\partial_{w}\phi-\partial_{u}\phi\partial_{w}^{2}\phi),\label{eq:SDG-eom}
\end{equation}
which we will use to define the correlators via multi-particle expansions. Note here that there is an explicit coordinate dependence ($u-v=2z$) on the right hand side, which requires a nontrivial ingredient in the multi-particle ansatz. 

Let us start with
\begin{align}
\phi(z)& =\sum_{P}e^{\mathrm{i}k_{P\mu}x^{\mu}} \Phi_P (z),\\
\Phi_{P}(z)&=\sum_{i=1}^{|P|}\sum_{P=P_{1}\cup\ldots\cup P_{i}}\sum_{n_{i}=0}\frac{1}{i!}\Phi^{(i)}(P_{1},\ldots,P_{i};n_{i})z^{n_{i}}e^{-E_{P_{1},\ldots,P_{i}}z}, \label{eq:SDG-ansatz}
\end{align}
where $P$ is now an ordered word, i.e., for $P=p_1 \ldots p_n$, we have $p_i < p_{i+1}$. The $i!$ factor removes the over counting introduced by the permutations of the words $\{P_1, \ldots ,P_i\}$ -- there is no way to distinguish them, unlike the group structure in SDYM. The single-particle solutions of \eqref{eq:SDG-eom} are given by $\Phi(z,k)=\exp (\mathrm{i}k_{p\mu}x^{\mu}-E_{p}z)$, with $\mu=t,x,y$. Furthermore, we have a polynomial dependence on the radial coordinate $z$ times the energy exponential. The degree of this polynomial is bounded, since each interaction vertex increases the power of $z$ by at most one, and it follows that
\begin{equation}
\Phi^{(i)}(P_{1},\ldots,P_{i};n_i\geq i)=0.
\end{equation}

The solution of the recursion follows the usual perturbiner approach. On the left hand side of equation \eqref{eq:SDG-eom}, we have
\begin{multline}
\Box_{\mathbb{R}^{4}}\phi=\sum_{P}e^{\mathrm{i}k_{P\mu}x^{\mu}}\sum_{i=1}^{|P|}\sum_{P=P_{1}\cup\ldots\cup P_{i}}\sum_{n_{i}=0}\frac{1}{i!}z^{n_{i}}e^{-E_{P_{1},\ldots,P_{i}}z}\\ \times\bigg(-S_{P_{1},\ldots,P_{i}}\Phi^{(i)}(P_{1},\ldots,P_{i};n_{i})-2(n_{i}+1)E_{P_{1},\ldots,P_{i}}\Phi^{(i)}(P_{1},\ldots,P_{i};n_{i}+1)\\+(n_{i}+2)(n_{i}+1)\Phi^{(i)}(P_{1},\ldots,P_{i};n_{i}+2)\bigg).
\end{multline}
As for the right hand side of \eqref{eq:SDG-eom}, we obtain
\begin{multline}
[\partial_{w}\phi\partial_{u}\partial_{w}\phi-\partial_{w}^{2}\phi\partial_{u}\phi]+z[\partial_{w}^{2}\phi\partial_{u}^{2}\phi-(\partial_{u}\partial_{w}\phi)^{2}]\\=\frac{1}{2}\sum_{P}e^{\mathrm{i}k_{P\mu}x^{\mu}}\sum_{i=2}^{|P|}\sum_{P=P_{1}\cup\ldots\cup P_{i}}\sum_{j=1}^{i-1}\sum_{n_{j}=0}\sum_{m_{i-j}=0}\left(\frac{1}{j!(i-j)!}\right)z^{n_{j}+m_{i-j}}e^{-E_{P_{1},\ldots,P_{i}}z}\\\times\bigg\{[(k_{P_{1},\ldots,P_{j}})_{w}-(k_{P_{j+1},\ldots,P_{i}})_{w}]X_{P_{1},\ldots,P_{j}|P_{j+1},\ldots,P_{i}}\\\times\Phi^{(j)}(P_{1},\ldots,P_{j};n_{j})\Phi^{(i-j)}(P_{j+1},\ldots,P_{i};m_{i-j})\\+(m_{i-j}+1)(k_{P_{1},\ldots,P_{j}})_{w}[(k_{P_{1},\ldots,P_{j}})_{w}-(k_{P_{j+1},\ldots,P_{i}})_{w}]\\\times\Phi^{(j)}(P_{1},\ldots,P_{j};n_{j})\Phi^{(i-j)}(P_{j+1},\ldots,P_{i};m_{i-j}+1)\\+z\bigg[-X_{P_{1},\ldots,P_{j}|P_{j+1},\ldots,P_{i}}^{2}\Phi^{(j)}(P_{1},\ldots,P_{j};n_{j})\Phi^{(i-j)}(P_{j+1},\ldots,P_{i};m_{i-j})\\-(m_{i-j}+1)(k_{P_{1},\ldots,P_{j}})_{w}X_{P_{1},\ldots,P_{j}|P_{j+1},\ldots,P_{i}}\\\times\Phi^{(j)}(P_{1},\ldots,P_{j};n_{j})\Phi^{(i-j)}(P_{j+1},\ldots,P_{i};m_{i-j}+1)\\+(n_{j}+1)(k_{P_{j+1},\ldots,P_{i}})_{w}X_{P_{1},\ldots,P_{j}|P_{j+1},\ldots,P_{i}}\\\times\Phi^{(j)}(P_{1},\ldots,P_{j};n_{j}+1)\Phi^{(i-j)}(P_{j+1},\ldots,P_{i};m_{i-j})\\-\frac{1}{2}(k_{P_{1},\ldots,P_{j}})_{w}^{2}(m_{i-j}+1)(m_{i-j}+2)\\\times\Phi^{(j)}(P_{1},\ldots,P_{j};n_{j})\Phi^{(i-j)}(P_{j+1},\ldots,P_{i};m_{i-j}+2)\\+\frac{1}{2}(k_{P_{1},\ldots,P_{j}})_{w}(k_{P_{j+1},\ldots,P_{i}})_{w}(n_{j}+1)(m_{i-j}+1)\\\times\Phi^{(j)}(P_{1},\ldots,P_{j};n_{j}+1)\Phi^{(i-j)}(P_{j+1},\ldots,P_{i};m_{i-j}+1)\bigg]\bigg\}.
\end{multline}

The multi-particle recursion is derived via the identification of the coefficients of $e^{\mathrm{i}k_{P\mu}x^{\mu}}e^{-E_{P_1,\ldots,P_i}z}z^{n_i}$ on both sides of the equation. As in the SDYM analysis, the coefficients $\Phi^{(1)}(P;0)$ parametrize free solutions. They are fixed by imposing the homogeneous Dirichlet boundary condition $\Phi_P(0)=0$ for $|P|>1$.

It is worth analysing some of the lower-point cases in order to understand the recursion above. For $P=12$, we obtain
\begin{multline}
    [S_{1,2}\Phi^{(2)}(1,2;0)+2E_{1,2}\Phi^{(2)}(1,2;1)+z\Phi^{(2)}(1,2;1)]e^{-E_{1,2}z}\\=[(k_{2w}-k_{1w})X_{1|2}+z\frac{1}{S_{1,2}}X_{1|2}^{2}]e^{-E_{1,2}z}.
\end{multline}
Therefore,
\begin{align}
S_{1,2}\Phi^{(2)}(1,2;0)&=(k_{2w}-k_{1w})X_{1|2}-2E_{1,2}\Phi^{(2)}(1,2;1),\\S_{1,2}\Phi^{(2)}(1,2;1)&=X_{1|2}^{2},
\end{align}
The two-particle function $\Phi_{12}(z)$ can then be written as
\begin{multline}
\Phi_{12}(z)\\=\Phi^{(1)}(12;0)e^{-E_{12}z}+\frac{1}{S_{1,2}}(k_{2w}-k_{1w})X_{1|2}e^{-E_{1,2}z}-X_{1|2}^{2}\frac{\partial}{\partial E_{1,2}}\left(\frac{1}{S_{1,2}}e^{-E_{1,2}z}\right).
\end{multline}
Note, in particular, that the linear dependence in $z$ is generated via a derivative of the energy exponential. By imposing $\Phi_{12}(0)=0$, the merged polarization is evaluated to be
\begin{equation}
\Phi^{(1)}(12;0)=\frac{1}{S_{1,2}}(k_{1w}-k_{2w})X_{1|2}+2X_{1|2}^{2}\frac{E_{1,2}}{S_{1,2}^{2}}.
\end{equation}
Higher-point examples follow the same logic but of course involve higher powers of $z$.

The $(N+1)$-point correlators of self-dual gravity are defined as
\begin{equation}
\mathcal{M}(1^+,\ldots, N^+,(N+1)^-) = \int_0^\infty  dz \left[ e^{-z E_{N+1}} (E_{1\ldots N}^2-\partial_z^2) \Phi_{1\ldots N}(z) \right].\label{eq:SDG-correlator}
\end{equation}
They are built in straight analogy with the SDYM case, though the explicit radial dependence of the interaction makes computations a bit more convoluted. Once more, it is worth looking at some lower-point examples. The three-point correlator is given by
\begin{equation}
\mathcal{M}(1^+,2^+,3^-)=\int_{0}^{\infty}dz\left[(k_{2w}-k_{1w})X_{1|2}+zX_{1|2}^{2}\right]e^{-E_{1,2,3}z}.
\end{equation}
The integration is straightforward to perform, yielding
\begin{equation}
\mathcal{M}(1^+,2^+,3^-)=X_{1|2}\left((k_{2w}-k_{1w})-X_{1|2}\frac{\partial}{\partial E_{1,2}}\right)\frac{1}{E_{1,2,3}},
\end{equation}
where we have used
\begin{equation}
\frac{\partial}{\partial E_{1,2}}\left(\frac{1}{E_{1,2,3}}\right)=-\left(\frac{1}{E_{1,2,3}}\right)^2.    
\end{equation}

For the four-point correlator, we obtain
\begin{multline}
\mathcal{M}(1^+,2^+,3^+,4^-)=\left((k_{2w}-k_{1w})X_{1|2}\frac{1}{S_{1,2}}-2X_{1|2}^{2}\frac{E_{1,2}}{S_{1,2}^{2}}\right)\left[\left(\frac{X_{1,2|3}}{E_{1,2,3,4}}\right)^{2}-\left(\frac{X_{12|3}}{E_{12,3,4}}\right)^{2}\right]\\+\left((k_{2w}-k_{1w})X_{1|2}\frac{1}{S_{1,2}}-2X_{1|2}^{2}\frac{E_{1,2}}{S_{1,2}^{2}}\right)(k_{3w}-k_{12w})\left[\left(\frac{X_{1,2|3}}{E_{1,2,3,4}}\right)-\left(\frac{X_{12|3}}{E_{12,3,4}}\right)\right]\\+2X_{1|2}^{2}X_{1,2|3}^{2}\frac{1}{S_{1,2}}\left(\frac{1}{E_{1,2,3,4}}\right)^{3}-k_{12w}X_{1|2}^{2}X_{1,2|3}\frac{1}{S_{1,2}}\left(\frac{1}{E_{1,2,3,4}}\right)^{2}\\-\frac{1}{2}(k_{3w}-k_{12w})k_{3w}X_{1|2}^{2}\frac{1}{S_{1,2}}\left(\frac{1}{E_{1,2,3,4}}\right)+\left[(12,3)\to\{(23,1),(13,2)\}\right].
\end{multline}
This expression has a simpler representation given by
\begin{multline}
\mathcal{M}(1^{+},2^{+},3^{+},4^{-})\\=\Phi^{(1)}(12;0)V_{12|3}\left(\frac{1}{E_{12,3,4}}\right)+V_{1|2}\left[\frac{1}{S_{1,2}}V_{1,2|3}\left(\frac{1}{E_{1,2,3,4}}\right)\right]\\+\Phi^{(1)}(13;0)V_{13|2}\left(\frac{1}{E_{13,2,4}}\right)+V_{1|3}\left[\frac{1}{S_{1,3}}V_{1,3|2}\left(\frac{1}{E_{1,2,3,4}}\right)\right]\\+\Phi^{(1)}(23;0)V_{23|1}\left(\frac{1}{E_{23,1,4}}\right)+V_{2|3}\left[\frac{1}{S_{2,3}}V_{2,3|1}\left(\frac{1}{E_{1,2,3,4}}\right)\right],\label{eq:SDG-4pt}
\end{multline}
with $\Phi^{(1)}(ij;0)$ denoting the merged polarizations. In particular, the first line on the right hand side corresponds to the following diagram,
\begin{eqnarray*}
&&
\begin{tikzpicture}[x=0.55cm,y=0.55cm, line cap=round, line join=round]
  \fill[gray!13] (0,0) circle (2.15);
    \draw[line width=0.2pt] (0,0) circle (2.15);

  \coordinate (p1) at (-1.95,0.98);
  \coordinate (p2) at (-1.95,-0.98);
  \coordinate (p5) at (0,2.15);
  \coordinate (p3) at (1.95,0.98);
  \coordinate (p4) at (1.95,-0.98);

  \coordinate (x1) at (-0.92,0);
  \coordinate (x2) at (0,0);
  \coordinate (x3) at (0.92,0);

  \draw[ dashed,line width=0.6pt] (p1) -- (x1);
  \draw[ dashed, line width=0.6pt] (p2) -- (x1);
  \draw[ dashed, line width=0.6pt] (x1) -- (x2) -- (x3);
  \draw[ dashed, line width=0.6pt] (p3) -- (x3);
  \draw[ dashed, line width=0.6pt] (p4) -- (x3);

  \fill (p1) circle (1.25pt);
  \fill (p2) circle (1.25pt);
  \fill (p3) circle (1.25pt);
  \fill (p4) circle (1.25pt);
  \fill (x1) circle (1.05pt);
  \fill (x3) circle (1.05pt);

  \node[font=\scriptsize, above left] at (-2.12,1.05) {$2^+$};
  \node[font=\scriptsize, below left] at (-2.12,-1.05) {$1^+$};
  \node[font=\scriptsize, above right] at (2.12,1.05) {$3^+$};
  \node[font=\scriptsize, below right] at (2.12,-1.05) {$4^-$};

  \node[font=\scriptsize, below] at (-0.92,-0.18) {$\,$};
  \node[font=\scriptsize, below] at (0,-0.18) {$\,$};
  \node[font=\scriptsize, below] at (0.92,-0.18) {$\,$};

  \node[font=\scriptsize, above] at (-0.46,0.12) {$\,$};
  \node[font=\scriptsize, above] at (0.46,0.12) {$\,$};
\end{tikzpicture}
\qquad
\makebox[206pt][r]{\raisebox{6.6ex}{$= \, V_{1|2}\, V_{1,2|3} \left(\frac{1}{S_{1,2} E_{1,2,3,4}} \right) - V_{1|2}\,V_{12|3} \left(\frac{1}{S_{1,2} E_{12,3,4}} \right).  $}} \nonumber
\end{eqnarray*}
Here we have defined the vertex operator
\begin{align}
    V_{P_{1},\ldots,P_{j}|P_{j+1},\ldots,P_{i}} & = X_{P_{1},\ldots,P_{j}|P_{j+1},\ldots,P_{i}} \mathcal{D}_{P_{1},\ldots,P_{j}|P_{j+1},\ldots,P_{i}},\\ \mathcal{D}_{P_{1},\ldots,P_{j}|P_{j+1},\ldots,P_{i}} & \equiv (k_{P_{j+1},\ldots,P_{i}})_{w}-(k_{P_{1},\ldots,P_{j}})_{w}-X_{P_{1},\ldots,P_{j}|P_{j+1},\ldots,P_{i}}\frac{\partial}{\partial E_{P_{1},\ldots P_{i}}},\label{eq:SDG-vertex}
\end{align}
such that
\begin{equation}
    \mathcal{D}_{P_{1},\ldots,P_{j}|P_{j+1},\ldots,P_{i}}=-\mathcal{D}_{P_{j+1},\ldots,P_{i}|P_{1},\ldots,P_{j}}.
\end{equation}
The expression for the 4-point correlator in \eqref{eq:SDG-4pt} is non-trivially equivalent to the one previously found in \cite{Chowdhury:2024dcy}, which we have also verified numerically. A key observation here is that the correlators of the the self-dual gravity theory can be obtained from those of the self-dual Yang-Mills theory via the insertion of differential operators in the interaction vertices, as shown in \eqref{eq:SDG-vertex}, and keeping track of operator ordering. This will be interpreted in terms of a double copy in the next section.

Following this cubic recursive structure, we can propose a compact formula for the correlator \eqref{eq:SDG-correlator}, given by
\begin{equation}
\mathcal{M}(1^{+},\ldots,N^{+},N+1)=\sum_{i=2}^{N}\sum_{1\ldots N=P_{1}\cup\ldots\cup P_{i}}\frac{1}{i!}\hat{M}(P_{1},\ldots,P_{i}|N+1)\left(\frac{1}{E_{P_{1},\ldots,P_{i},N+1}}\right),\label{eq:SDG-new-correlator}
\end{equation}
where $\hat{M}(P_{1},\ldots,P_{i}|N+1)$ can be thought of as an \textit{amplitude operator} acting on the energy poles, expressed as
\begin{multline}
\hat{M}(P_{1},\ldots,P_{i}|N+1)=\frac{1}{2}\sum_{j=1}^{i-1}\frac{1}{j!(i-j)!}\Gamma_{P_{1},\ldots,P_{j}}\Gamma_{P_{j+1},\ldots,P_{i}}V_{P_{1},\ldots,P_{j}|P_{j+1},\ldots,P_{i}}\\+\textrm{perm. }\{P_{1,}\ldots,P_{i}\}.
\end{multline}
The multi-particle operators $\Gamma_{P_1,\ldots,P_i}$ are recursively determined via the relation
\begin{multline}
\Gamma_{P_{1},\ldots,P_{i}}=\frac{1}{2}\sum_{j=1}^{i-1}\frac{1}{j!(i-j)!}\Gamma_{P_{1},\ldots,P_{j}}\Gamma_{P_{j+1},\ldots,P_{i}}\left[V_{P_{1},\ldots,P_{j}|P_{j+1},\ldots,P_{i}}\left(\frac{1}{S_{P_{1},\ldots,P_{i}}}\right)\right]\\+\textrm{perm. }\{P_{1},\ldots,P_{i}\}.
\end{multline}
Single-word operators $\Gamma_P$ are given by the merged polarizations $\Phi^{(1)}(P;0)$, which are expressed as
\begin{equation}
\Gamma_{P}=-\sum_{i=2}^{|P|}\sum_{P=P_{1}\cup\ldots\cup P_{i}}\frac{1}{i!}\Gamma_{P_{1},\ldots,P_{i}}.
\end{equation}
For one-letter words we simply have $\Gamma_p=1$. Note that the vertex operators act on the generalized Mandelstam variables, energy poles, as well as on $X_{P_1,\ldots,P_j|P_{j+1},\ldots,P_i}$ given in \eqref{Xs}. For instance, it is easy to show that
\begin{equation}
\frac{\partial X_{P_{1},\ldots P_{j}|P_{j+1},\ldots,P_{i}}}{\partial E_{P_{1}}}=\frac{1}{2}(k_{P_{j+1},\ldots,P_{i}})_{w},
\end{equation}
and this identity holds for any derivative with respect to energy variables with labels in $\{P_1,\ldots,P_j\}$. 

Just like in the self-dual Yang--Mills case, the construction of higher points correlators, including the identification of specific diagram contributions, is performed with almost no effort. For example, the five-point contributions coming from the analogue SDG diagrams of equations \eqref{eq:SDYM-5pt-12345} and \eqref{eq:SDYM-5pt-34512} are respectively given by
\begin{multline*}
    V_{1|2} \Bigg\{V_{1,2|3}V_{1,2,3|4} \left(\frac{1}{S_{1,2}S_{1,2,3}E_{1,2,3,4,5}}\right) - V_{12|3}V_{12,3|4}\left(\frac{1}{S_{1,2}S_{12,3}E_{12,3,4,5}}\right)\\- V_{1,2|3} V_{123|4} \left(\frac{1}{S_{1,2}S_{1,2,3}E_{123,4,5}}\right)
+ V_{12|3} V_{123|4} \left(\frac{1}{S_{1,2}S_{12,3}E_{123,4,5}}\right) \Bigg\},
\end{multline*}
and
\begin{multline*}
V_{1|2}V_{3|4} \Bigg\{V_{1,2|3,4} \left(\frac{1}{S_{1,2}S_{3,4}E_{1,2,3,4,5}}\right) - V_{12|3,4}\left(\frac{1}{S_{1,2}S_{3,4}E_{12,3,4,5}}\right)\\
- V_{1,2|34} \left(\frac{1}{S_{1,2}S_{3,4}E_{1,2,34,5}}\right)
+ V_{12|34} \left(\frac{1}{S_{1,2}S_{3,4}E_{12,34,5}}\right)\Bigg\}.
\end{multline*}
Here, however, we have to be careful about the ordering of the operators, which typically do not commute.

In general, the analysis of the flat space limit is also straightforward, coming from the highest pole of the total energy. For the three-point correlator we have
\begin{equation}
\left.\mathcal{M}(1^{+},2^{+},3^{-})\right|_{\textrm{flat}}\to\left(\frac{X_{1|2}}{E_{1,2,3}}\right)^{2},   
\end{equation}
while the four-point flat space limit can be expressed as
\begin{equation}
\left.\mathcal{M}(1^{+},2^{+},3^{+},4^{-})\right|_{\textrm{flat}}\\\to2\left[\frac{X_{1|2}^{2}X_{1,2|3}^{2}}{S_{1,2}}+\frac{X_{1|3}^{2}X_{1,3|2}^{2}}{S_{1,3}}+\frac{X_{2|3}^{2}X_{2,3|1}^{2}}{S_{2,3}}\right]\left(\frac{1}{E_{1,2,3,4}}\right)^{3},
\end{equation}
The term inside the square brackets on the right hand side vanishes using momentum conservation, i.e., $E_{1,2,3,4}=0$, as it corresponds to the four-point SDG flat space amplitude. At five points, we have numerically verified that our proposal does not have the spurious poles apparently introduced by the generalized Mandelstam variables, and that it yields the correct flat space limit. The latter for higher point correlators is straightforward to obtain analytically, since we must look at the residue of the higher order total energy pole. For an $(N+1)$-point correlator, the dominant total energy pole is $(E_{1,\ldots,N,N+1})^{-N}$, and its residue is proportional to the on-shell vanishing flat space amplitude.

\section{Color-kinematics duality} \label{colorkin}

The cubic structure of the self-dual Yang--Mills theory is very suggestive, aso we will now analyze it in the context of color-kinematics duality \cite{Bern:2008qj,Bern:2019prr}. For example, the four-point amplitudes with merged external polarizations in equation \eqref{eq:SDYM-flat-amplitude} can be recast as
\begin{equation}
A(P_{1},P_{2},P_{3}|N+1)\\=\Phi^{(1)}(P_{1})\Phi^{(1)}(P_{2})\Phi^{(1)}(P_{3}) \left(\frac{N_{[[P_{1},P_{2}],P_{3}]}}{S_{P_{1},P_{2}}}-\frac{N_{[[P_{2},P_{3}],P_{1}]}}{S_{P_{2},p_{3}}}\right),
\end{equation}
where we identify the numerator
\begin{equation}
    N_{[[P_{i},P_{j}],P_{k}]}\equiv X_{P_{i},P_{j}|P_{k}}X_{P_{i}|P_{j}}.
\end{equation}
We are using the standard bracket notation for binary trees (see e.g. \cite{Mafra:2020qst}), which is graphically expressed as
\begin{equation*}
  [[P_i,P_j],P_k]
  \;\longrightarrow\;
  \begin{tikzpicture}[baseline={(0,0.4)}, scale=0.8, line width=0.6pt]
    \node[anchor=base] at (0,1.25) {$P_i$};
    \node[anchor=base] at (1,1.25) {$P_j$};
    \node[anchor=base] at (2,1.25) {$P_k$};
    \draw (0,1)   -- (0.5,0.5);
    \draw (1,1)   -- (0.5,0.5);
    \draw (0.5,0.5) -- (1,0);
    \draw (2,1)   -- (1,0);
    \draw (1,0)   -- (1,-0.4);
  \end{tikzpicture}.
\end{equation*}
Since $X_{P_i|P_j}=-X_{P_j|P_i}$, it is easy to show that they satisfy the kinematic Jacobi identity
\begin{multline}
N_{[[P_{i},P_{j}],P_{k}]}+N_{[[P_{j},P_{k}],P_{i}]}+N_{[[P_{k},P_{i}],P_{j}]}\\=X_{P_{i}|P_{k}}X_{P_{i}|P_{j}}+X_{P_{j}|P_{i}}X_{P_{j}|P_{k}}+X_{P_{k}|P_{j}}X_{P_{k}|P_{i}}\\+X_{P_{j}|P_{k}}X_{P_{i}|P_{j}}+X_{P_{k}|P_{i}}X_{P_{j}|P_{k}}+X_{P_{i}|P_{j}}X_{P_{k}|P_{i}}=0.\label{eq:SDYM-BCJ}
\end{multline}
This is a consequence of the cubic character of the SDYM interactions, tied to the Jacobi identity of the color factors.

The construction for higher points is straightforward to generalize (e.g. \cite{Edison:2020ehu,Mafra:2020qst, Frost:2020eoa}). We define the master BCJ numerators as
\begin{equation}
N_{\ell[P_{1},P_{2}, \cdots ,P_{k}] }= X_{P_1|P_2}X_{P_1,P_2|P_3}\ldots X_{P_1,P_2,\cdots,P_{k-1}|P_k },
\end{equation}
where $\ell$ denotes the complete left bracketing,
\begin{equation}
\ell\left[P_{1},P_{2}, \cdots ,P_{k}\right]= [[\ldots[P_1,P_2],P_3],\ldots, P_k] .
\end{equation} 
They encode $(k-1)!$ crossing symmetric numerators, satisfying the Jacobi identity by construction. The master numerators can be used to generate the complete amplitude with color-kinematics duality manifest. For instance, one possible contribution to the four-point amplitude is expressed as
\begin{equation}
N_{[[P_{1},P_{2}],[P_{3},P_{4}]]}=N_{\ell[P_{1},P_{2},P_{3},P_{4}]}-N_{\ell[P_{1},P_{2},P_{4},P_{3}]}.
\end{equation}
Since the amplitudes \eqref{eq:SDYM-flat-amplitude} are the building blocks in \eqref{eq:SDYM-Nptcorrelator}, our representation of color-ordered SDYM correlators makes color-kinematics duality manifest for each of the energy poles, where the merged polarizations are treated as independent objects.

As for the SDG correlators, we immediately observe a double copy structure at the level of the (cubic) interaction vertices, with
\begin{equation}
V_{P_{1},\ldots,P_{j}|P_{j+1},\ldots,P_{i}}=\underbrace{X_{P_{1},\ldots,P_{j}|P_{j+1},\ldots,P_{i}}}_{\textrm{SDYM}}\times \, \mathcal{D}_{P_{1},\ldots,P_{j}|P_{j+1},\ldots,P_{i}}.  
\end{equation}
Note, for instance, that we can define the differential four-point numerator
\begin{equation}
\hat{N}_{[[P_{i},P_{j}],P_{k}]} = \mathcal{D}_{P_{i}|P_{j}}\mathcal{D}_{P_{i},P_{j}|P_{k}},
\end{equation}satisfying
\begin{equation}
\left(\hat{N}_{[[P_{i},P_{j}],P_{k}]}+\hat{N}_{[[P_{j},P_{k}],P_{i}]}+\hat{N}_{[[P_{k},P_{i}],P_{j}]}\right) f(E_{P_i,P_j,P_k})=0,    
\end{equation}
which is analogous to the Jacobi identity in equation \eqref{eq:SDYM-BCJ}. In fact, as long as we are careful with operator ordering, higher-point correlators in SDG are reproduced via a double-copy procedure. This can be seen already in equation \eqref{eq:SDG-4pt}. The double-copy is trivial in the three-point energy poles ($E_{12,3,4}$, $E_{13,2,4}$, and $E_{23,1,4}$), with merged polarizations. As for the contributions to the total energy pole ($E_{1,2,3,4}$), they can be cast as
\begin{multline}
\hat{M}(1,2,3|4)\left( 
\frac{1}{E_{1,2,3,4}}
\right)=\left[\hat{N}_{[[1,2],3]}N_{[[1,2],3]}\left(\frac{1}{S_{1,2}}\right)+\hat{N}_{[[2,3],1]}N_{[[2,3],1]}\left(\frac{1}{S_{2,3}}\right)\right.\\+\left.\hat{N}_{[[1,3],2]}N_{[[1,3],2]}\left(\frac{1}{S_{1,3}}\right)\right]\left(\frac{1}{E_{1,2,3,4}}\right),
\end{multline}
where the differential numerators also act on the SDYM numerators and on the generalized Mandelstam variables. More generally, a four-point contribution to the correlator will take the form
\begin{multline}
\hat{M}(P_1,P_2,P_3|N+1)\left( 
\frac{1}{E_{P_1,P_2,P_3,N+1}}
\right)\\
=
\Gamma_{P_1}\Gamma_{P_2}\Gamma_{P_3}
\left[\hat{N}_{[[P_1,P_2],P_3]}N_{[[P_1,P_2],P_3]}\left(\frac{1}{S_{P_1,P_2}}\right)+\hat{N}_{[[P_2,P_3],P_1]}N_{[[P_2,P_3],P_1]}\left(\frac{1}{S_{P_2,P_3}}\right)\right.\\+\left.\hat{N}_{[[P_1,P_3],P_2]}N_{[[P_1,P_3],P_2]}\left(\frac{1}{S_{P_1,P_3}}\right)\right]\frac{1}{E_{P_1,P_2,P_3,N+1}}.
\end{multline}

As an additional test, we have verified the double copy construction of the SDG correlator at five points via a numerical comparison with the diagrammatic computation.

\section{Conclusion} \label{conclusion}

In this paper, we have presented all-multiplicity formulae for boundary correlators of scalar theories which describe the self-dual sectors of Yang-Mills and gravity in AdS$_4$. These formulae take the form of an amplitude expansion similar to the one recently derived for full Yang-Mills theory in \cite{Gomez:2026yno}. In particular, an $n$-point correlator is written as a sum over flat space amplitudes of multiplicity $\leq n$ weighted by corresponding energy poles, where the amplitudes are computed using BG recursion without energy conservation and the external data of the lower-point amplitudes is obtained by merging the external data of higher-point amplitudes. In the SDG case, the amplitudes appearing in the expansion are differential operators which can be obtained from the SDYM amplitudes using a simple double copy prescription. This can in turn be understood from an underlying color-kinematics duality in terms of (differential) kinematic numerators, which holds at arbitrary multiplicity in the amplitude expansion due to the fact that amplitudes are naturally expressed in terms of cubic graphs.   

There are a number of directions to further explore. In the first instance it would be interesting to uplift these results to spinning correlators. It was recently pointed out that the $\theta$-term and Neumann boundary conditions play an important role in spinning SDYM correlators \cite{Skvortsov:2026gtq,Skvortsov:2026ofl,Skvortsov:2026dru,Heuveline:2026knw}, so it would be interesting to incorporate boundary terms and non-Dirichlet boundary conditions in our construction. On the other hand, our results can be thought of as a toy model for Yang-Mills and Einstein gravity correlators with standard Dirichlet boundary conditions. In particular, our results suggest that an amplitude expansion should exist for Einstein gravity correlators which can be uplifted from the Yang-Mills formula recently obtained in \cite{Gomez:2026yno} using a double copy prescription where kinematic numerators are promoted to differential operators. Color-kinematics duality and the double copy in (A)dS$_4$ have recently been explored from various perspectives \cite{Farrow:2018yni,Armstrong:2020woi,Albayrak:2020fyp,Jain:2021qcl,Diwakar:2021juk,Herderschee:2022ntr,Lee:2022fgr, Mei:2023jkb,Bala:2026sfh, Bala:2026jsj}, and we believe that the amplitude expansion may be useful for this purpose because it allows for the importation of useful results from flat space. 

There has recently been progress in extending the $w_{1+\infty}$ symmetry of asymptotically flat gravity \cite{Strominger:2021mtt,Ball:2021tmb,Monteiro:2022lwm,Adamo:2021lrv,Miller:2025wpq,Guevara:2025tsm} to (A)dS$_4$ background \cite{Lipstein:2023pih,Taylor:2023ajd,Bittleston:2024rqe,DiGiacomo:2026oku}, with important connections to three-dimensional conformal field theory \cite{Strominger:2026yrh,Himwich:2026exq,Zhu:2026ova}. It would therefore be particularly interesting to see how such symmetries are realized by the correlators derived in this paper. It would also be interesting to explore alternative geometric representations of these correlators based on cosmological polytopes \cite{Arkani-Hamed:2017fdk,Arkani-Hamed:2024jbp}, differential equations \cite{Arkani-Hamed:2023bsv}, scattering equations \cite{Roehrig:2020kck,Eberhardt:2020ewh,Gomez:2021qfd}, or the orthogonal Grassmannian \cite{Arundine:2026fbr}, which have played an important role in the study of flat space scattering amplitudes. Finally, it would be intriguing to extend our all-multiplicity formulae to loop-level. In flat space, loop amplitudes in the self-dual sector are rational functions at one-loop and vanish at higher loops \cite{Bern:1993qk,Bern:1996ja,Bern:1998xc,Bern:1998sv}, so we expect to see similar elements of simplicity in (A)dS background. 

In summary, the study of correlation functions in (A)dS background is entering an exciting new phase with the recent discovery of several all-multiplicity formulae \cite{Gomez:2021qfd,Gomez:2026yno,Mathys:2026pdq,Heuveline:2026knw,He:2026sja,Nabavi:2026zyk}, analogous to the Parke-Taylor formula in flat space \cite{Parke:1986gb}. We expect that this will lead to important new insights into the mathematical structure of quantum field theory in curved spacetime with applications to holography and cosmology, and that the self-dual sector should play an important role in these developments. 

\begin{acknowledgments}

We thank Chandramouli Chowdhury and Silvia Nagy for useful discussions. H.G. and  R.L.J. are supported by the GA\v{C}R grant 25-16244S from the Czech Science Foundation. 
HG would like to thank partial support from FAPESP, grant 2026/00820-8. A.L. is supported by an STFC Consolidated Grant UKRI4226.

\end{acknowledgments}

\bibliographystyle{JHEP}
\bibliography{references}

\end{document}